\documentclass[aps,reprint,pre]{revtex4-1}
\usepackage[T1]{fontenc}
\usepackage{amstext}
\usepackage{amssymb}
\usepackage{graphicx}

\makeatletter
\@ifundefined{textcolor}{}
{%
 \definecolor{BLACK}{gray}{0}
 \definecolor{WHITE}{gray}{1}
 \definecolor{RED}{rgb}{1,0,0}
 \definecolor{GREEN}{rgb}{0,1,0}
 \definecolor{BLUE}{rgb}{0,0,1}
 \definecolor{CYAN}{cmyk}{1,0,0,0}
 \definecolor{MAGENTA}{cmyk}{0,1,0,0}
 \definecolor{YELLOW}{cmyk}{0,0,1,0}
 }

\makeatother

\begin{document}

\title{Quantifying the complexity of random Boolean networks}

\author{Xinwei Gong}

\author{Joshua E.~S.~Socolar}

\affiliation{Center for Nonlinear and Complex Systems and Physics Department,
Duke University, Durham, NC, USA}
\begin{abstract}
	 	 	
We study two measures of the complexity of heterogeneous extended systems, taking random Boolean networks as prototypical cases. A measure defined by Shalizi et al. for cellular automata, based on a criterion for optimal statistical prediction  [Shalizi et al., Phys. Rev. Lett. 93, 118701 (2004)], does not distinguish between the spatial inhomogeneity of the ordered phase and the dynamical inhomogeneity of the disordered phase. A modification in which complexities of individual nodes are calculated yields vanishing complexity values for networks in the ordered and critical regimes and for highly disordered networks, peaking somewhere in the disordered regime. Individual nodes with high complexity are the ones that pass the most information from the past to the future, a quantity that depends in a nontrivial way on both the Boolean function of a given node and its location within the network.
\end{abstract}
\maketitle

\section{Introduction}

%Traditional information-theoretic definitions of complexity measure the diversity of a system's temporal patterns, assigning higher complexity values to systems whose time series require longer descriptions. By these measures, periodic behaviors have low complexity and random processes have high complexity.  If a statistical description is considered sufficient, however, random processes can be described very simply. For example, the series of heads and tails generated by tossing a coin can be described succinctly as ``a Bernoulli process with probability 0.5.''
 
In computational mechanics, the complexity of a process that generates a single time series is defined as the least amount
of information required for a maximally accurate statistical description of the series
\cite{springerlink:10.1007/BF00668821,PhysRevLett.63.105,springerlink:10.1023/A:1010388907793}.
This definition classifies random processes, as well as simple periodic ones,
as having low complexity.  In a 2004 article, Shalizi et al.\
extended the definition to spatially--extended dynamical systems and introduced
an algorithm for measuring the complexity of a discrete
system given time series data for all components \cite{PhysRevLett.93.118701}.
Applying the algorithm to 2D cyclic cellular automata
(CCA) confirmed that it classified CCA rules generating fixed states or incoherent
local oscillations as having low complexity and cases that produce
turbulent spiral waves as having high complexity.

Shalizi's complexity measure, $C_{\mu}$, is defined as the amount of information
stored in local causal states, where a causal state is an
equivalence class of all past configurations that give rise to the same distribution of future outcomes.  A set of causal states can be discerned from time series data for all elements in the system.  
$C_{\mu}$ is obtained by measuring correlations between truncated past and future light cones associated with each element in the system.

The formalism developed by Shalizi et al.\ is intriguing due to its principled application of information theoretic concepts to the analysis of complex dynamical systems.  It is not immediately clear, however, how the formalism can be applied to random or heterogeneous networks.  We show here that a natural extension requires distinguishing between averages over nodes and averages over time, and also that the relevant quantities can be computed analytically for certain classes of random Boolean networks.  Our primary goal is to clarify the meaning of Shalizi's complexity measure.  In doing so, we propose a new measure of the information processing occurring at a given node, which gives a new statistic for distinguishing the roles played by different nodes in a complex network.  We show that this measure depends upon global features of the dynamics and, in particular, that it is not simply related to the sensitivity of the network and is not maximized in critical networks.  Along the way, we also present a conjecture concerning the dynamics of Boolean networks consisting only of parity functions.

We study two complexity measures that differ only in the choice of the ensembles of spacetime points used for averaging the local complexity.
One approach considers the ensemble
of all spatial points at the same time instant, which corresponds to Shalizi's $C_{\mu}$.
In computing $C_{\mu}$, all nodes are given equal weight in determining the probabilities
of observing different states.  Shalizi et al.\ used $C_{\mu}$ to investigate self-organization
of cellular automata, which are logically and topologically uniform
networks of discrete, interacting elements.  

A second approach is to assign a complexity to each individual element by averaging over time.  The system average
of these individual complexities is denoted $C_{\nu}$.
For a homogeneous system in which all elements have statistically indistinguishable time series (a regular lattice with rules that lead to turbulent dynamics, for example), $C_{\mu}$ and $C_{\nu}$ are the same.
We suggest that $C_{\nu}$ is the one that is most informative for spatially inhomogeneous systems.

To develop conceptual insights into the behavior of the quantities $C_{\mu}$ and $C_{\nu}$, we study their dependence on the parameters specifying different ensembles of random Boolean
networks (RBNs).  RBNs were first studied by Kauffman as toy models of gene regulatory networks \cite{SA1969437,citeulike:312346}.
They have garnered much attention in the last few decades,
and features such as steady-state bias, sensitivity
to perturbations, attractor length and mutual information between
nodes have been extensively investigated \cite{0295-5075-1-2-001,PhysRevE.74.046104,PhysRevLett.94.088701,PhysRevE.77.011901}. 
We show here that for any given distribution of logic functions, the value of $C_{\mu}$ for a RBN can be analytically calculated as a function of the bias $\rho$ and is not simply related to the sensitivity $\lambda$. 
$C_{\nu}$ on the other hand, is always near zero for sensitivity
values $\lambda\le1$, where the network is in the ordered or critical
regime, and is also zero for the highest possible $\lambda$
value, where the network dynamics is strongly chaotic.  Thus $C_{\nu}$ reflects the intuitive notion that systems with short periodic cycles or apparently random behavior should both have low complexity.  
The maximum of $C_{\nu}$ for RBNs occurs somewhere in the disordered regime. 
We find also
that the amount of information processed by any given individual depends
on global properties of the network dynamics as well as the logic functions of the node in question and others in its neighborhood.

Section II defines the two measures
$C_{\mu}$ and $C_{\nu}$ in detail. Section III describes an implementation of these definitions in the context of RBNs and presents theoretical and numerical results on the complexity of a certain class of RBNs.
The relation between network complexity and sensitivity, and the relation between individual nodes' complexity and role in determining the network dynamics is also discussed.
We conclude with some general remarks and suggestions for future research.

\section{Complexity Measures}

The Grassberger-Crutchfield-Young statistical complexity is defined as the least
amount of information about the past trajectory required for optimal
prediction of future trajectory, given
time series data for a single variable~\cite{springerlink:10.1023/A:1010388907793}. This measure
is calculated from time series data alone, without reference to the physical laws that govern the dynamical processes.
Shalizi et al.\ extended the concept to processes with spatial extent~\cite{PhysRevLett.93.118701}. We summarize the basic ideas here.
For details and information--theoretic support of these ideas, see
Ref.~\cite{DMTCS-AB0102}.

Given a field $\textit{X}$ that varies over space and time in a system where
information propagates at a maximum speed of $\textit{c}$, the past
light cone of a space-time point $(\vec{r},t)$ consists of all space-time
points where events can influence $X(\vec{r},t)$. 
The future light cone consists of all points that can be influenced by $X(\vec{r},t)$.
$L^{-}(\vec{r},t)$ and $L^{+}(\vec{r},t)$
denote the configurations of the field $\textit{X}$ in the past and futre light cones, respectively:
\begin{eqnarray}
L^{\pm}(\vec{r},t) &  = & \{\left(X(\vec{s},u),|t-u|,\vec{r}-\vec{s}\right) \nonumber \\
\ & \ & \quad\forall\; u\gtrless t {\rm\ and\ } |\vec{s}-\vec{r}|\leqslant c|t-u|\}.
\end{eqnarray}
Note that each element of $L^{\pm}$ consists of three quantities: a field
value, a time lapse, and a {\em relative} position. Thus $L^{-}$ (or $L^{+}$) is exactly the same for two distinct space-time points with identical past (or future) light cones.

Each space-time
point is associated with one $L^{-}$ and one $L^{+}$.  
An ensemble of such pairs defines
a probability distribution $P(L^{+}|L^{-})$ for future light cone configurations conditioned
on past configurations.
A causal state $\epsilon(L^{-})$ is defined as a set of past light cone configurations that have the
same distribution of future configurations.  All instances
of a given causal state predict the same distribution of future light
cone configurations:
\begin{equation}
\ensuremath{\epsilon(l^{-})=\text{\{}\omega:P(L^{\text{+}}|L^{-}=\omega)=P(L^{\text{+}}|L^{-}=l^{-})\text{\}}.}
\label{eq:causalstates}
\end{equation}

By definition, $\epsilon(l^{-})$ is a sufficient local statistic;
knowing the causal state at a given point provides the same predictive power
as knowing the exact past light cone configuration.
$\epsilon(l^{-})$ is also a {\it minimal} sufficient statistic~\cite{kullback1997information},
meaning that the sufficient statistic $\epsilon(l^{-})$ contains the least
amount of information among all statistics that have the same predictive
power:
\begin{equation}
\ensuremath{H[\epsilon(l^{-})]\leqslant H[\eta(l^{-})]},
\end{equation}
where $\eta(l^{-})$ is a sufficient statistic and $H[X]=-\sum_{i}P(X=x_{i})\log_{2}P(X=x_{i})$
denotes Shannon entropy.  It then follows that $H[\epsilon(l^{-})]$
is the least amount of information for optimal prediction of the future dynamics~\cite{DMTCS-AB0102, PhysRevLett.93.118701}, which is taken to be the relevant measure of a system's complexity.  We use the shorthand notation
\begin{equation}
\ensuremath{C\equiv H[\epsilon(l^{-})]}.
\end{equation}

The value of $C$ depends upon the choice
of the ensemble of space-time points used to determine the causal
states.  We consider two choices.
For the first, denoted $C_{\mu}$ and studied by Shalizi et al.\ for homogeneous systems~\cite{PhysRevLett.93.118701, PhysRevE.73.036104}, causal
states are determined at any given time by considering the ensemble
of spatial locations in the system at that time. In other words, causal states $\epsilon_{\mu}(l^{-},\; t)$ are defined as in Eq.~\ref{eq:causalstates} with the restriction that $P(L^{\text{+}}|L^{-})$ is determined from the set of past and future light cone pairs present at time $t$.
This approach allows one to speak of the complexity of a system as
a function of time, which may exhibit transient dynamics.  
Systems that exhibit a spontaneous increase
in $C_{\mu}(t)$ have been described by Shalizi et al.\ as going through
a self-organization process \cite{PhysRevLett.93.118701}.
For present purposes, we use
$C_{\mu}$ to refer to the complexity after transients have decayed.

Alternatively, for $C_{\nu}$ the causal states at a given spatial point are defined using the ensemble of light cone pairs observed at different times. In other words, causal states $\epsilon_{\nu}(l^{-},\; \vec{r})$ are defined as in Eq.~\ref{eq:causalstates} with the restriction that $P(L^{\text{+}}|L^{-})$ is determined from the set of past and future light cones pairs occurring at different times at $\vec{r}$. 
$C_{\nu}(\vec{r})$ receives negligible weight from transients, and in practice we
compute it by taking data only after transients have relaxed.  
Because $C_{\nu}(\vec{r})$ is associated with a particular element of the system, it
provides information about the role that element plays in the dynamical evolution
of the system.
Averaging over all $\vec{r}$, we obtain
a global complexity measure $C_{\nu}$.

In practice, estimating the complexity requires restricting the depths of the light cone configurations to a manageable size.
For the Boolean networks discussed below, we will show
that it is sufficient to consider a single update step in the past
and a single step in the future.

\section{Complexity of Random Boolean Networks}

We study $C_{\mu}(t)$ and $C_{\nu}(r)$ in
synchronously updated RBNs to determine how (or whether) they
are related to well-understood measures of the dynamics, such as
the sensitivity of the network or the overall bias in the values of the binary variables.
In a synchronous RBN, at each time step each 
node $i$ is updated based on a logic function $f_i$ that is applied to the current values
of its input nodes:
\begin{equation}
\ensuremath{x_{i}(t)=f_{i}(x_{i_1}(t-1),x_{i_2}(t-1),\dots,x_{i_n}(t-1)),}\label{eq:RBN}
\end{equation}
where $t$ is a positive integer.
Here $f_{i}:\text{\{0,1\}}^{n}\rightarrow\text{\{0,1\}}$ is a quenched
Boolean logic function for node $\textit{i}$, chosen randomly from some a weighted distribution over
the logic functions with the appropriate number of inputs, and the quenched choice of 
nodes $i_j$ that act as inputs to $i$ are randomly selected, independently for each $i$, from the full set of nodes, with each given equal weight.
The $x_{i_j}$'s are the binary values of the inputs to $i$.
An important feature of RBN's is that the density of feedback or feedforward loops of any finite size goes to zero as the network size goes to infinity~\cite{alon2007introduction}; the local structure around each node is ``tree-like.'' 

The number
of inputs and outputs per node, which may or may not be the same for
all nodes, and the distribution of logic functions are the parameters
that characterize an ensemble of synchronous RBNs. From these parameters,
two global measures, the bias and the sensitivity, can
be calculated analytically~\cite{0295-5075-1-2-001, PhysRevE.74.046104}. The bias $\rho$ is the fraction of nodes with
value 1 after transients have decayed. Let $\textbf{x}\in\text{\{0,1\}}^{K}$ be an input vector of length $K$, and $K$ be the fixed number of input per node for a network, then the bias map of the network is given by
\begin{equation}
\ensuremath{\rho(t+1)=\big{<}{\displaystyle\sum_{\textbf{x}}}f(\textbf{x})\rho(t)^{|\textbf{x}|}(1-\rho(t))^{K-|\textbf{x}|}\big{>},}\label{eq:bias}
\end{equation}
where $\left\langle \bullet\right\rangle $ denotes expectation taken over distribution of logic functions and $|\textbf{x}|$ is the number of 1s in $\textbf{x}$. The fixed point or cyclic behavior of the bias is determined by the bias map. The sensitivity $\lambda$ is the average rate of increase of the Hamming distance between
two state space trajectories that initially differ at a single node:
\begin{equation}
\ensuremath{\lambda=\big{<}{{\displaystyle\sum_{i=1}^{K}}}{\displaystyle\sum_{\textbf{x}}}(f (\textbf {x}^{(i, 0)})\oplus f (\textbf {x}^{(i, 1)}))\rho^{|\textbf{x}|}(1-\rho)^{K-|\textbf{x}|}\big{>},}
\label{eq:sensitivity}
\end{equation}
where $\textbf{x}^{(i,j)}=(x_{1},...,x_{i-1},j,x_{i+1},...,x_{K})$ and $\oplus$ denotes the XOR function \cite{PhysRevE.74.046104}. The sensitivity distinguishes qualitatively different network behaviors
that have been termed ordered ($\lambda<1$), disordered ($\lambda>1$), and
critical ($\lambda = 1$)~\cite{2002nlin......4062K}.

For the present study, we fix the number of inputs and outputs of the nodes such that
the light cones of all nodes have the same shape.  We study the simplest nontrivial case,
in which all nodes have exactly two inputs and two outputs.  For these networks,
the greatest possible sensitivity is $\lambda = 2$.  

Ref.~\cite{PhysRevLett.93.118701} outlined an algorithm for computationally
distinguishing the causal states of cellular automata, and we can
implement similar procedures to distinguish the causal states of our RBNs.
Because nodes in RBNs are not assigned spatial positions, the definition
of $L^{-}$ and $L^{\text{+}}$ requires some technical modifications.
In the calculation of $C_{\mu}$, $L_{\mu}^{-}(i,t)$ of node $\textit{i}$
at time step $\textit{t}$ is defined as
\begin{eqnarray}
\ensuremath{L_{\mu}^{-}(i,t) & = & \text{\{}\text{(}X(j,u),\; t-u,\; d(i,j)\text{)} \nonumber \\
\ & \ & \quad\forall\; u<t\text{ and }d(i,j)\le t-u\text{\}}},
\end{eqnarray}
where $\textit{d(i, j) }$ is the shortest distance between nodes
$\textit{i}$ and $\textit{j}$  (the minimum number of links that must be traversed to get from node $\textit{i}$ to node $\textit{j}$), and the speed of propagation of information is now 1. By this definition, different nodes
that have the same distance from a given node are deemed indistinguishable
for the purpose of identifying a light cone configuration. 
Although the update rules (Eq.~\ref{eq:RBN})
contain indices for node inputs, the calculation
of statistical complexity is defined without reference to the underlying physical laws and hence without the knowledge of which input is which at any given node.

For calculating the complexity $C_{\nu}(i)$ of an individual node,
however, different inputs and outputs are distinguishable.
When comparing two past or future light cone configurations of the same node
\emph{i} at different time steps, one can keep track of which input is which and observe that when
the two inputs differ, the
future light cone configuration depends upon which input value is 1.
Consequently, $L_{\nu}^{-}(i,t)$ is defined as
\begin{equation}
\ensuremath{L_{\nu}^{-}(i,t)=\text{\{}\text{(}X(j,u),\; t-u,\; j\text{)}\;\forall\; u<t\text{ and }d(i,j)\le t-u\text{\}}}.
\end{equation}

\begin{figure*}
\includegraphics[scale=0.65]{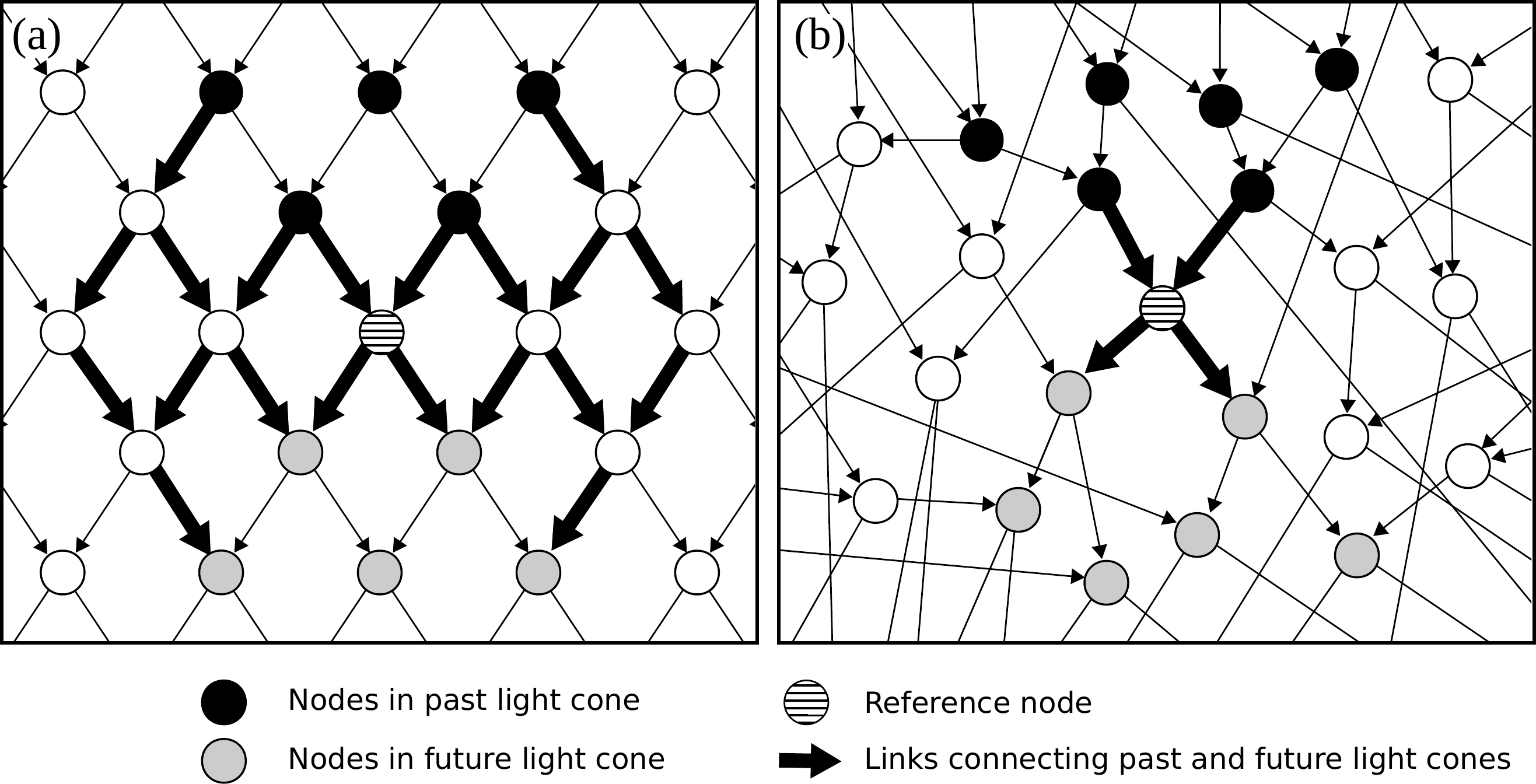}
\caption{Local portions of (a) a two-dimensional regular lattice network and (b) a random network. Black/grey nodes are in the past/future light cones that are truncated at a depth of two time steps. The light cone configuration includes the value of each node at a time corresponding to the number of links that must be crossed to reach the reference node. In (a), a node's past light cone influences its future light cone through multiple paths. In (b) the past light cone influences the future light cone only through the reference node itself.}
\label{fig:rbn}
\end{figure*}

In measuring $C_{\mu}$ or $C_{\nu}$ for a RBN in the limit of large system size, as opposed to a regular lattice, it is sufficient to restrict the computation to light cones of depth one.
The the lack of memory in the update rules ensures that all the relevant information for determining the state of a node is determined by the configuration one time step in the past, and the lack of short loops in the random graph ensures that all information propagating from a given past light cone to a given future light cone must pass through the single node under consideration.
Fig.~\ref{fig:rbn} illustrates the latter point.
In a regular lattice, a node's
past light cone can influence its future light cone through multiple
paths,
whereas in an RBN, the chance of finding such a path is vanishingly small in the limit of large
system size.  Thus,
even though the definition of statistical complexity relies on arbitrarily
deep light cones, the results for depth 1 become exact
for RBNs with system size $N\rightarrow\infty$.  This fact
enables us to calculate $C_{\mu}$ and, in some cases, $C_{\nu}$ analytically in the large system limit.

$C_{\mu}$ is calculated from depth-1 light cones as follows.   
For a node with two inputs $\textit{i}$ and
$\textit{j}$, there are only three possible past light cone configurations:
$\{0, 0\}$, $\{1,1\}$, and $\{1,0\}$ or $\{0,1\}$ (note we have omitted the relative time and distance entries in writing the light cone configurations because they are now trivial after we restrict the light cone depth to 1), occurring with probability $(1-\rho)^{2}$, $\rho^{2}$, and $2\rho(1-\rho)$,
respectively, and yielding probabilities for the reference node being
ON of $\langle f(0,0)\rangle$, $\langle f(1,1)\rangle$, and $1/2(\langle f(0,1)\rangle+\langle f(1,0)\rangle)$ respectively. As discussed above, the probability of observing
a future light cone configuration, $L^{+}$, depends only on the state
of the reference node. Thus, if each
of the three possible $L^{-}$'s yields a unique probability for
the reference node to be ON or OFF, which in turn yields a unique distribution of $L^{+}$'s, then each $L^{-}$ is itself a causal state and we have
\begin{eqnarray}
\ensuremath{C_{\mu} & = & -\rho^{2}\log_{2}(\rho^{2})-(1-\rho)^{2}\log_{2}((1-\rho)^{2}) \nonumber \\
\ & \ & \quad -2\rho(1-\rho)\log_{2}(2\rho(1-\rho))}\,.\label{eq:cmu}
\end{eqnarray}

Modifications to Eq.~\ref{eq:cmu} are required if the above assumptions do not hold.  
For example, if any two
of $\langle f(0,0)\rangle$, $1/2(\langle f(0,1)\rangle+\langle f(1,0)\rangle)$
and $\langle f(1,1)\rangle$ are equal, then there would be less than three causal states.  
The following scenario is also possible.
If the state of a reference node $x$ is $1$, the probability that an
output is 1 is 
\begin{equation}
\rho_{1}^{+}=\rho\left\langle f(1,1)\right\rangle +1/2(1-\rho)\left\langle f(1,0)\right\rangle +1/2(1-\rho)\left\langle f(0,1)\right\rangle\,;
\end{equation}
and if $x=0$, the probability that an output is 1 is 
\begin{equation}
\rho_{0}^{+}=(1-\rho)\left\langle f(0,0)\right\rangle +1/2\rho\left\langle f(1,0)\right\rangle +1/2\rho\left\langle f(0,1)\right\rangle\,.
\end{equation}
In the case of an accidental degeneracy $\rho_{1}^{+}=\rho_{0}^{+}$, the distribution
of $L^{+}$ is independent of $x$ and therefore independent of $L^{-}$, in which case all three possible
$L^{-}$ collapse to a single causal state, yielding $C_{\mu}=0$.

\begin{figure*}
\includegraphics[scale=0.75]{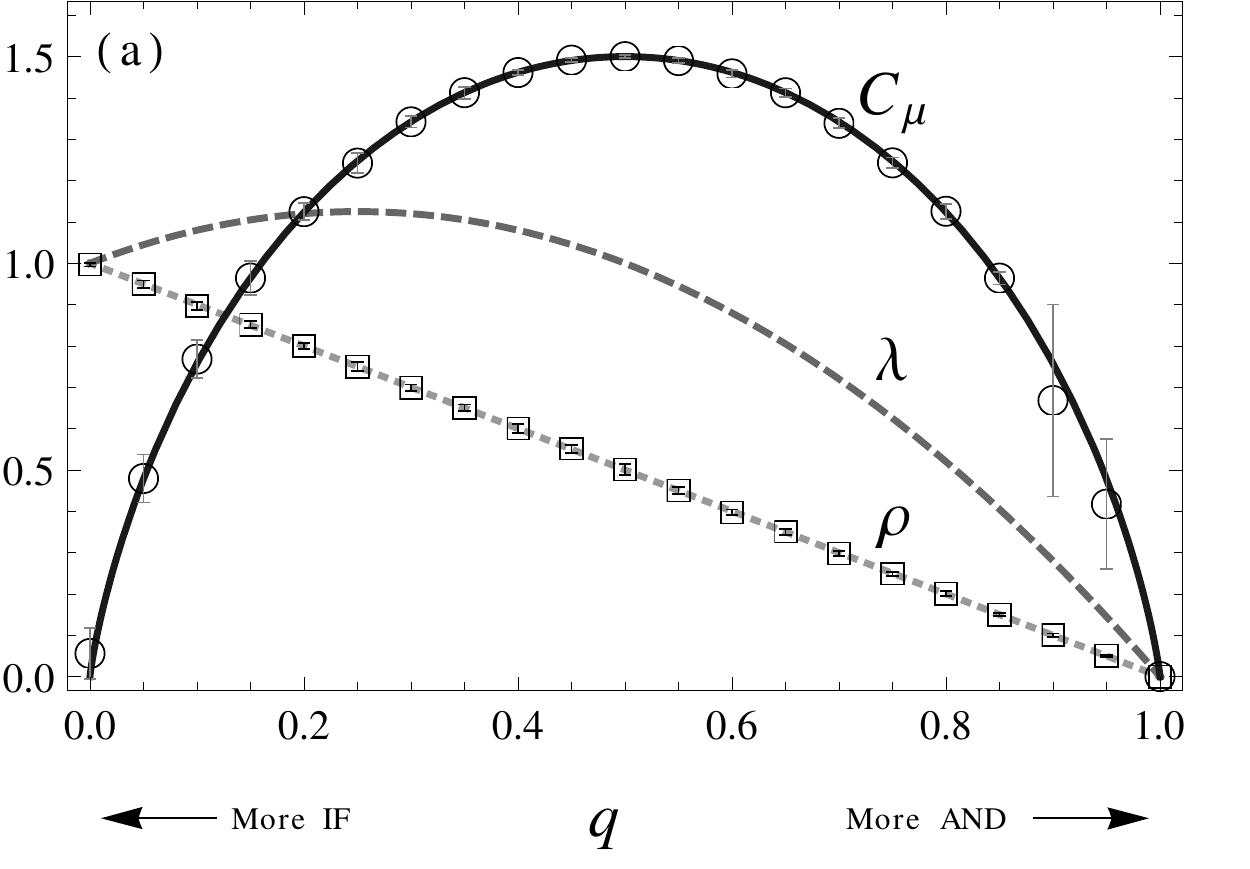}\includegraphics[scale=0.75]{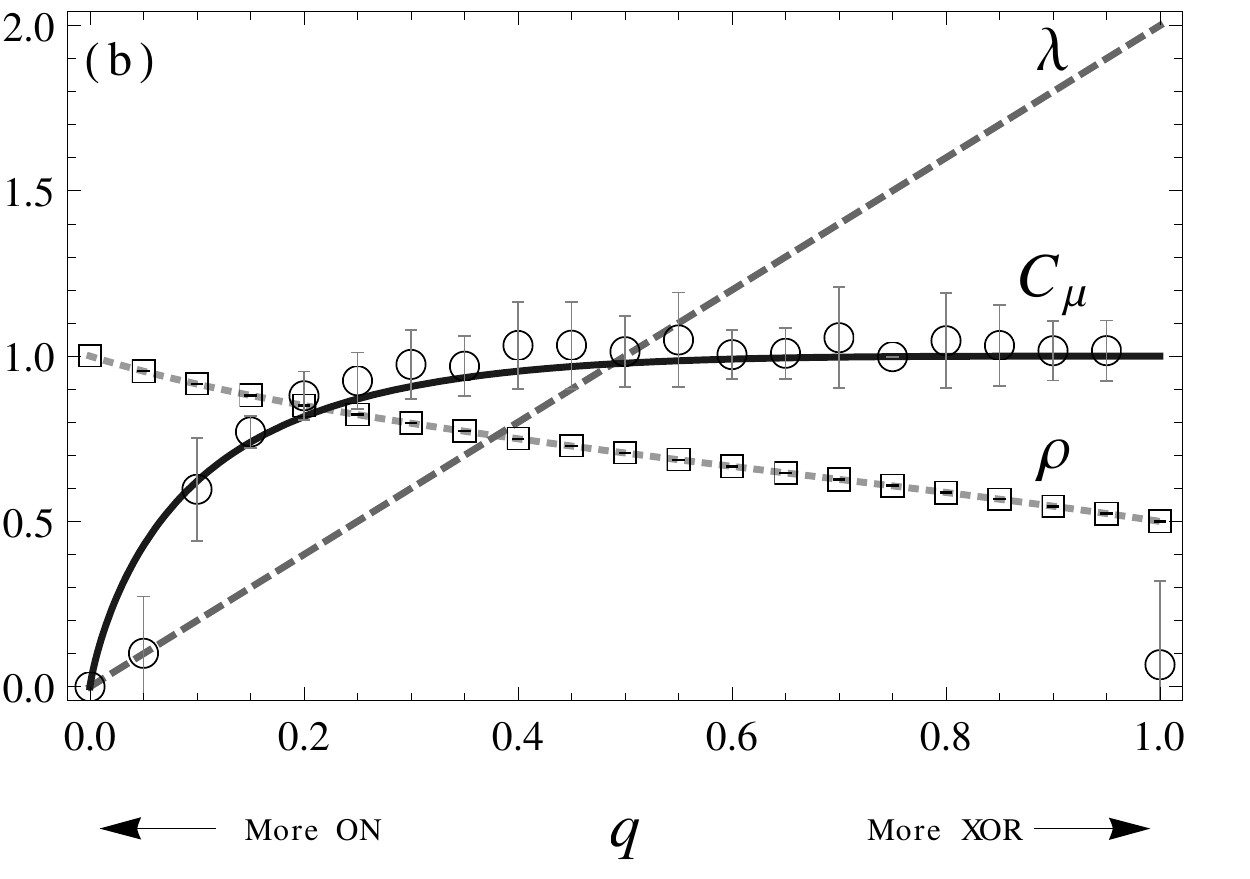}
\caption{The steady-state bias $\rho$, sensitivity $\lambda$ and complexity $C_{\mu}$ for networks consisting of two types of nodes.
(a) A fraction $q$ of the nodes are assigned the AND function $(f(0,0),f(0,1),f(1,0),f(1,1)) = (0,0,0,1)$ while the others are assigned $(1,0,1,1)$ (IF).  (b) A fraction $q$ are assigned $(0,1,1,0)$ (XOR) and the rest $(1,1,1,1)$ (always ON). The solid curve and circular points are respectively the analytical and simulation results for $C_{\mu}$. The size of networks is $N=10^{4}$,
and time of data collection for each network realization is $T=10^{6}$. For each of the 21 ratios of the two logic functions (21 dots in the graph), 30 RBNs are constructed. For each RBN, 30 runs are simulated with different initial conditions. }
\label{fig:Cmu}
\end{figure*}

Fig.~\ref{fig:Cmu}
shows the comparison between the analytical calculation of $C_{\mu}$
and simulation results for ensembles of networks with mixtures of
two logic functions.  The horizontal axis denotes the fraction of
nodes that get assigned the indicated Boolean function. The ensemble in Fig.~\ref{fig:Cmu}(a) satisfies the requirements for Eq.~\ref{eq:cmu}. The ensemble in Fig.~\ref{fig:Cmu}(b) only has two causal states because $\langle f(0,0)\rangle$=$\langle f(1,1)\rangle$=$1-q$, and $1/2(\langle f(0,1)\rangle+\langle f(1,0)\rangle)$=1, where $q$ is the fraction of XOR nodes in a XOR-ON network. The first two terms in Eq. \ref{eq:cmu} collaps into one for such an ensemble, and the corresponding $C_{\mu}$ is given by
\begin{eqnarray} \ensuremath{C_{\mu}&=&-(\rho^{2}+(1-\rho)^2)\text{log}_{2}(\rho^{2}+(1-\rho)^2) \nonumber \\
&&-2\rho(1-\rho)\text{log}_{2}(2\rho(1-\rho))}.
\end{eqnarray}
From Eq. \ref{eq:bias} we can obtain the fixed-state bias for a XOR-ON network to be 
\begin{equation}
\ensuremath{\rho=\frac{-1+2q+\sqrt{1+4q-4q^{2}}}{4q}},
\end{equation}
which is monotonically decreasing from $\rho=1$ to $\rho=0.5$ for $q\in[0,1]$. We could subsequently obtain a complicated expression of $C_{\mu}$ in terms of $q$ which we would omit here, but we can see that $C_{\mu}(q)$ is an increasing function in $q\in[0,1]$ because $C_{\mu}(\rho)$ monotonically decreases in $\rho\in[0.5,1]$. 
Simulation results indicate that $C_{\mu}$ does not change noticeably for a given  ensemble for network sizes above $N=1000$. 
The simulations for $N=10^4$
agree well with the large system analytical result.  The disagreements at $q=0.05$ and $q=1$ in the Fig.~\ref{fig:Cmu}b 
are due, respectively, to the fact that differences between causal states are too small for the simulations to resolve ($1-q$ too close to $q$) and to
a collapse of the type described in the previous paragraph ($\rho_{1}^{+}=\rho_{0}^{+}$). For some choices of Boolean functions, the bias $\rho$ can oscillate, which leads to persistent oscillations in $C_{\mu}$. 

Critical networks (with $\lambda = 1$) have been hypothesized to have properties that might be favored by natural selection or other self-organized processes~\cite{citeulike:312346}.
Our findings show that $C_{\mu}$ is not simply related to sensitivity, so that
maximization of $C_{\mu}$ does not correspond to selection of critical networks.
Fig.~\ref{fig:Cmu} illustrates this point with two
examples.  In (a), we see that the network is critical at both $q=0$ and $q=0.5$ and that $C_{\mu}$ is a maximum in one case but zero in the other.  In (b), we see that $C_{\mu}$ increases monotonically as $\lambda$ increases from 1 (the critical value) to 2.

The fact that $C_{\mu}$ can be high in ordered systems ($\lambda < 1$), where the attractor dynamics is trivial, highlights the fact that the spatial inhomogeneity of states alone can produce a high complexity.
In these networks, almost all nodes are frozen on a fixed value, independent of the initial conditions, but different nodes may be frozen on different values and $C_{\mu}$ becomes a measure of the degree of variation from node to node.
The fact that $C_{\mu}$ can be high in strongly disordered systems ($\lambda$ near the maximum possible value of 2),
reflects the tendency of the bias $\rho$ to approach values that maximize $C_{\mu}$ at these points.

The calculation of $C_{\nu}$ treats each component
as an agent with its own causal states and complexity, then averages those complexity values.  When causal states are determined by considering
the past and future light cones of a single node at different times, every frozen
node has zero complexity, as does any node for which the past and future
light cones are uncorrelated.  In ordered or critical networks, where
only a vanishingly small number of nodes are not frozen, $C_{\nu}$
is very close to zero.  In highly disordered systems, the behavior of most nodes closely approximates a purely stochastic process, so $C_{\nu}$ is again near zero.  Thus $C_{\nu}$ is maximized somewhere in the disordered regime.

\begin{figure}[b]
\includegraphics[scale=0.75]{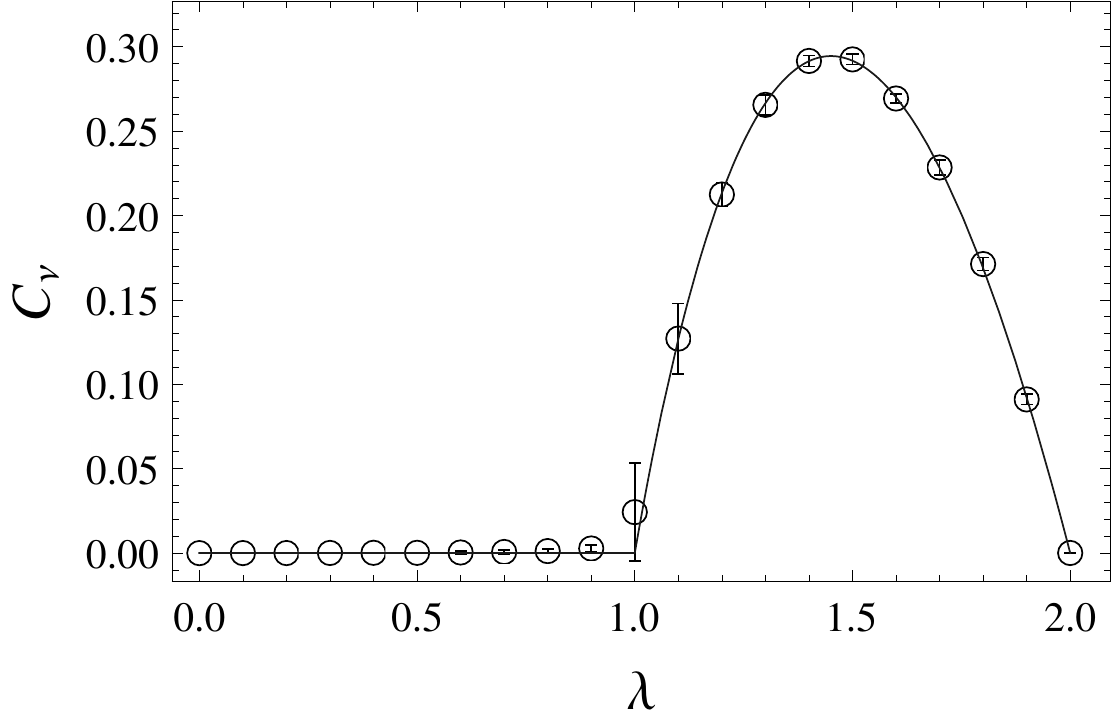}
\caption{A typical pattern for complexity $C_{\nu}$ vs sensitivity $\lambda$.  $C_{\nu}$ is close to zero for $\lambda\le 1$ and for the maximum value of $\lambda=2$. $C_{\nu}$ is maximized near $\lambda=1.5$.  The two logic functions are $(0,1,1,0)$ and $(1,1,1,1)$, as in Fig.~\ref{fig:Cmu}(b). The solid curve shows theoretical results and the circular dots show simulation results.  The size of networks is $N=10^{4}$ and time of data collection for each network realization is $T=10^{6}$.
For each of the 21 ratios of the two logic functions (21 dots in the
graph), 30 RBNs are constructed. For each RBN, 30 runs are simulated
with different initial conditions.}
\label{fig:Cnu}
\end{figure}

Fig.~\ref{fig:Cnu} shows a typical plot of
complexity $C_{\nu}$ as a function of sensitivity $\lambda$.  Recall that $\lambda=2$
is the highest sensitivity value possible for a system in which each node has exactly two outputs.
All ensembles of systems with a full range
of sensitivity values exhibit the same general relation between $C_{\nu}$ and $\lambda$,
with $C_{\nu}$ maximized at different $\lambda$ from ensemble to ensemble.

We have not found a way to calculate $C_{\nu}$ analytically for networks with a general combination of logic functions, but we can do it (in the large system limit) for the special case of the networks represented in Fig.~\ref{fig:Cmu}(b), which contain only the logic functions XOR (0110) and ON (1111).  
In this case, the complexity of each node is either 0 or 1, depending on whether
it and/or its neighbors are frozen or not.  

The first step in calculating $C_{\nu}$ is to find the fraction $\gamma$ of nodes that are frozen.  In general, for large networks of two-input logic functions in which the probability that a node is frozen when a subset of its inputs are frozen is independent of the value of those frozen inputs, $\gamma$ is given by a solution of the equation
\begin{equation} \label{eqn:samuelsson}
p_0 (1-x)^2 + 2 p_1 x(1-x) + p_2 x^2 = x\,,
\end{equation}
where $p_k$ is the probability that a node will be frozen if exactly $k$ of its inputs are frozen.  (See~\cite{PhysRevE.74.036113}, Eq.~(70).) The XOR-ON networks satisfy the requirements and have $p_0 = p_1 = 1-q$ and $p_2 = 1$, where $q$ is the fraction of XOR nodes, which yields 
\begin{equation} \label{eqn:gammaq}
\gamma = (1-q)/q {\rm \ or\ }1\,.
\end{equation}  
For $q>1/2$, $\gamma=(1-q)/q$ is the stable solution. For $q<1/2$, $\gamma=1$ is the stable solution, implying that the fraction of frozen nodes approaches unity.

The unfrozen nodes are all XORs, some of which have one frozen input and therefore act either as a copier or inverter of the other input.  The network of unfrozen nodes thus acts as a network in which every node executes a parity function (or its inversion). We propose the following general conjecture about such networks.

{\bf Conjecture:} {\em Every node in a synchronously updated Boolean network containing only parity functions (for arbitrary numbers of inputs to each node) will have bias of exactly $1/2$ when averaged over trajectories originating from all possible initial states of the network, where the bias of an individual node at a given time $t$ is defined as the fraction of trajectories for which that a node is ON at time $t$.  Moreover, the bias is not affected if the nodes executing parity functions are embedded in a larger network in which all other nodes freeze after some transient.}  

The conjecture is supported by numerical simulations of 10,000 randomly generated 16-node networks with only XOR and ON gates.  We find that the bias of exactly $1/2$ on each unfrozen node is maintained on each individual time step.  For many of these networks, it can be shown that every state of the unfrozen portion has exactly one pre-image, which is enough to guarantee that all states occur with equal frequency when averaged over initial conditions.  In cases where some  states have two pre-images, the number of recurrent states decreases by a factor of two (and may in some cases be reduced by additional factors of 2).  

Our conjecture provides the basis for a computation of $C_{\nu}$.
First note that in the large system limit, all frozen nodes have complexity $C_{\nu}(i)=0$. 
For an unfrozen node in the XOR-ON network, $C_{\nu}(i)$ may be 0, which occurs when an unfrozen XOR has outputs only to unfrozen nodes that all have additional unfrozen inputs.  The value of each output node in this case is equally likely to take either value, no matter what the value of node $i$.  $C_{\nu}(i)$ may be nonzero, however, if the unfrozen node has at least one output to a node whose other input is frozen.  The future light cone distribution then depends on $x_i$, and the past light cones that yield the two values occur with equal probability, which gives $C_{\nu}(i)= 1$.

Calculating $C_{\nu}$ for the XOR-ON ensemble of networks comes down to determining the fraction of nodes that satisfy the condition for having $C_{\nu}(i)=1$, which can be obtained from a mean-field calculation as follows. Let $X_{1}$ and $X_{2}$ be inputs of node $X_{3}$ and assume that $X_1$ has no other output. If we assume that $X_{1}$ is an unfrozen node, then the probability that $X_{2}$ is frozen and $X_{3}$ is unfrozen is $q\gamma$. Because the only way for $X_1$ to have complexity 1 is for $X_2$ to be frozen, the probability that $X_1$ has complexity $1$ is $q\gamma$.  For an unfrozen node with two outputs, the complexity will be $1$ if either of the outputs has a frozen input, which occurs with probability $1-(1-q\gamma)^2$.  Thus the $C_{\nu}$ is equal to the total fraction of nodes with nonzero complexity:
\begin{equation}
C_{\nu} = (1-\gamma)(1-(1-q\gamma)^2)\,.
\end{equation}
Using Eq.~\ref{eqn:gammaq}, we have the system average
\begin{equation} \label{eqn:Cnuq}
C_{\nu} = \frac{1}{q}(1-q^2)(2q-1)\,.
\end{equation}
Eq.~\ref{eqn:Cnuq} agrees well with simulation results (Fig.~\ref{fig:Cnu}).  

The XOR-ON example shows that $C_{\nu}$ can be calculated for specific distributions of logic functions and, more importantly, illustrates that the complexity of an individual component depends on globally determined dynamics, not just on the logical process carried out by the individual node or on the smaller network motifs containing the node.  The pattern of frozen nodes is generated by a transient process that may propagate through the entire network \cite{PhysRevE.74.036113}.

The precise import of the value of $C_{\nu}(i)$ at a given node is not immediately clear.
We have measured the correlation between $C_{\nu}(i)$ and a new measure $\delta(i)$ that characterizes the effect of replacing node $i$ with a random number generator.  $\delta(i)$ is determined as follows.  Two realizations of the same network are run in parallel with the same initial condition.
After running long enough for transients to decay, the  
logic function $f_{i}$ is ignored in one of the copies and node $i$ is replaced by a stochastic agent generating $x_{i}=1$ or $0$ with probabilities $p$ and $1-p$, respectively, on each time step.  
Over the course of $M$ steps, we calculate
the fraction of time that each node in the network is ON for each of the realizations, forming two
$N$-dimensional vectors.
The Euclidean distance between these
two vectors is denoted $\delta(i,p)$, and we define $\delta(i)\equiv\min\{\delta(i,p):p\in[0,1]\}$.
We obtain an approximate measure of $\delta(i)$ by taking $M=10 N$ and considering $p = 0.1\,n$ for $n = 0,1,\ldots,10$.  
A study of the XOR-AND and IF-AND ensembles shows that the correlation coefficient between $C_{\nu}(i)$ and $\delta(i)$ in the disordered regime
ranges between 0.4 to 0.7 for network size N=1000.
This suggests an interpretation of $C_{\nu}(i)$: it is a measure of importance of the logical processing performed at node $i$ for  determining the global dynamics.  A low value of $C_{\nu}(i)$ means that the computations done by node $i$ can be effectively simulated by a random number generator.  

\begin{figure}
\includegraphics[width=0.85\columnwidth]{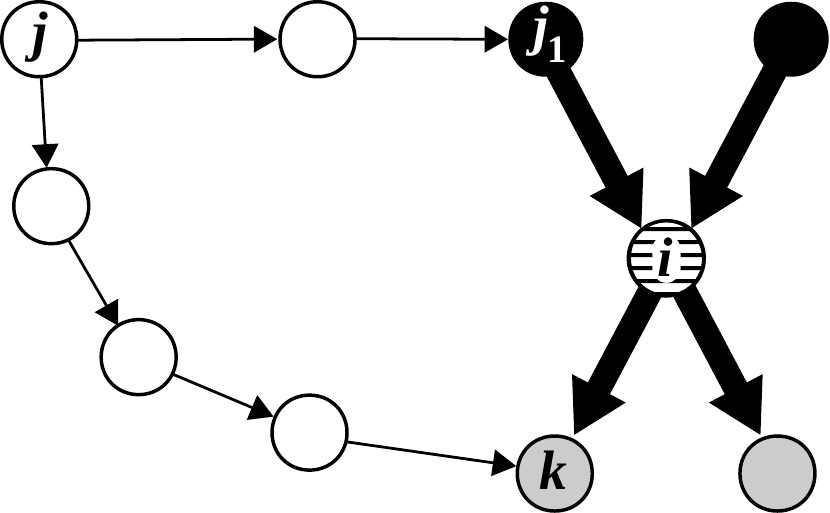}
\caption{An example of a network structure that can cause a frozen nodes to have nonzero $C_{\nu}(i)$: node $i$ is frozen, but its past and future light cones are correlated because nodes $j_{1}$ and $k$ receive input from the same node $j$ through chains of unfrozen nodes that differ in length by exactly two links.}
\label{fig:twopaths}
\end{figure}

We note that correlations associated with the existence of two paths in the network from one node to another can cause frozen nodes to have nonzero $C_{\nu}(i)$.  The correlation arises when a path of unfrozen nodes from some node $j$ passes through the frozen node $i$ and then immediately to an unfrozen node $k$, while another path of unfrozen nodes of the same length goes from $j$ to $k$ without passing through $i$.  Let $j_1$ be the input to $i$ along the first path.  Then $x_{j1}$ and $x_{k}$ may be correlated due to the common source at node $j$, in spite of the fact that node $i$ does not pass on any information. (See Fig.~\ref{fig:twopaths}.) In the large system limit, the fraction of nodes affected by this type of correlation vanishes.

Because $\delta(i)$ is necessarily zero for any frozen node, the existence of multiple paths in a finite system causes some nodes with high $C_{\nu}(i)$ to have low $\delta(i)$.  We further note that $\delta(i)$ itself is not a perfect measure of dynamical importance of a given node's activity because the average Euclidean distance between the bias vectors may be small even though the sequence of states is substantially different.  In fact, we observe that $\delta(i)$ saturates at low $C_{\nu}(i)$ values.  Comparing $C_{\nu}(i)$ with more precise measures of dynamical importance would be an interesting topic for future research. 

\section{Conclusion}

In extending the formalism of Shalizi's complexity measure to random Boolean networks,
a distinction must be made between determining causal states by averaging over nodes at a given time step versus averaging
over time for each node separately.  The two methods are equivalent for systems described by
the same input-output function at every node of a uniform lattice, but yield different results
for systems that are spatially inhomogeneous either because the input-output functions are
different for different nodes or because the network topology is not a regular lattice.
The networks we have studied have both types of inhomogeneity.

We find that $C_{\mu}$, obtained from statistics at a single time step,
can be calculated analytically for RBN's, and that it is not simply related to sensitivity
to small perturbations.  The maximum of $C_{\mu}$ can occur in the ordered,
disordered, or critical regime, depending on the details of the probability distribution chosen
for the Boolean rules in the network.  $C_{\mu}$ is directly related to the steady state
(or long-term oscillatory) bias in the node values.

$C_{\nu}$, obtained from statistics compiled over time for individual nodes,
is zero (in the large system limit) everywhere in the ordered and at the limiting
value of sensitivity in the disordered regime, so it is maximized somewhere in the disordered regime.
The value at an individual node, $C_{\nu}(i)$, may be interpreted as a measure
of the role that node in determining the global dynamics.  Nodes that can be
replaced by random number generators without substantially altering the global dynamics
(and hence perform no essential information processing) tend to have lower $C_{\nu}(i)$.
This last finding may be helpful in characterizing the behavior of social or biological systems
and the individual agents or components that comprise them.
Interestingly, the $C_{\nu}(i)$'s depend on the global features of the network that determine
its attractors, not just on the local logic functions and topology.  
In other words, the identification of important players in a network requires global information about the network, not just a characterization of each individual's behavior.

\begin{acknowledgments}
We thank C.~R.~Shalizi for conversations in the early stages of this work, and X.~Cheng and M.~Sun for helpful discussions. 
\end{acknowledgments}

\end{document}